\documentclass[pre,twocolumn,aps,showpacs]{revtex4-1}

\usepackage{amsmath}
\usepackage{amssymb}
\usepackage{amsbsy}
\usepackage{graphicx}
\usepackage{color}
\usepackage{pdfpages}

\newcommand{\moy}[1]{\left\langle #1 \right\rangle}

\newcommand{\ex}[1]{\mathrm{e}^{#1}}

\newcommand{\emuu}[0]{\mathbf{e_{\boldsymbol{\mu}}}}
\newcommand{\lamb}[0]{\boldsymbol{\lambda}}
\newcommand{\enu}[0]{\mathbf{e_{\boldsymbol \nu}}}

\newcommand{\zz}[0]{\mathbf{0}}

\begin{document}

\title{Microscopic theory for negative differential mobility in  crowded environments}

\author{O. B\'enichou}
\affiliation{Sorbonne Universit\'es, UPMC Univ Paris 06, UMR 7600, LPTMC, F-75005, Paris, France}
\affiliation{CNRS, UMR 7600, Laboratoire de Physique Th\'{e}orique de la Mati\`{e}re Condens\'{e}e, F-75005, Paris, France}

\author{P. Illien}
\affiliation{Sorbonne Universit\'es, UPMC Univ Paris 06, UMR 7600, LPTMC, F-75005, Paris, France}
\affiliation{CNRS, UMR 7600, Laboratoire de Physique Th\'{e}orique de la Mati\`{e}re Condens\'{e}e, F-75005, Paris, France}

\author{G. Oshanin}
\affiliation{Sorbonne Universit\'es, UPMC Univ Paris 06, UMR 7600, LPTMC, F-75005, Paris, France}
\affiliation{CNRS, UMR 7600, Laboratoire de Physique Th\'{e}orique de la Mati\`{e}re Condens\'{e}e, F-75005, Paris, France}

\author{A. Sarracino}
\affiliation{Sorbonne Universit\'es, UPMC Univ Paris 06, UMR 7600, LPTMC, F-75005, Paris, France}
\affiliation{CNRS, UMR 7600, Laboratoire de Physique Th\'{e}orique de la Mati\`{e}re Condens\'{e}e, F-75005, Paris, France}

\author{R. Voituriez}
\affiliation{Sorbonne Universit\'es, UPMC Univ Paris 06, UMR 7600, LPTMC, F-75005, Paris, France}
\affiliation{CNRS, UMR 7600, Laboratoire de Physique Th\'{e}orique de la Mati\`{e}re Condens\'{e}e, F-75005, Paris, France}

\begin{abstract}
  We study the behavior of the stationary velocity of a driven
  particle in an environment of mobile hard-core obstacles.  Based on
  a lattice gas model, we demonstrate analytically that the drift
  velocity can exhibit a nonmonotonic dependence on the applied force,
  and show quantitatively that such negative differential mobility
  (NDM), observed in various physical contexts, is controlled by both
  the density and diffusion time scale of obstacles.  Our study
  unifies recent numerical and analytical results obtained in specific
  regimes, and makes it possible to determine analytically the region
  of the full parameter space where NDM occurs. These results suggest
  that NDM could be a generic feature of biased (or active) transport
  in crowded environments.
\end{abstract}


\maketitle

\emph{Introduction.}--Quantifying the response of a complex system to
an external force is one of the cornerstone problems of statistical
mechanics.  In the linear response regime, a fundamental result is the
fluctuation-dissipation theorem, which relates system response and
spontaneous fluctuations.  Within the last years a great effort has
been devoted to generalizations of this theorem to nonequilibrium
situations~\cite{MPRV08,C11,seifert,gradenigo2}, when the time
reversal symmetry is broken, and also to elucidating the effects of
the higher order contributions in the external
perturbation~\cite{M86,BB05,LCSZ08,D12,ben1,ben2,ben3}. From
experimental perspective, theoretical understanding of the latter
issues is of an utmost importance in several fields, such as active
microrheology~\cite{weeks,SM09,poon} and dynamics of nonequilibrium
fluids~\cite{EM08,WB09}.

A striking example of anomalous behavior beyond the linear regime is
the negative response of a particle's velocity to an applied force,
observed in diverse situations in which a particle subject to an
external force $F$ travels through a medium.  The terminal drift
velocity $V(F)$ attained by the driven particle is then a nonmonotonic
function of the force: upon a gradual increase of $F$, the terminal
drift velocity first grows as expected from linear response, reaches a
peak value and eventually decreases.  This means that the differential
mobility of the driven particle becomes negative for $F$ exceeding a
certain threshold value.  Such a counter-intuitive ``getting more from
pushing less''~\cite{royce} behavior of the differential mobility (or
of the differential conductivity) has been observed for a variety of
physical systems and processes, e.g. for electron transfer in
semiconductors at low temperatures~\cite{conwell,NCCGO76,SBW86,LHC91},
hopping processes in disordered media \cite{bryksin}, transport of
electrons in mixtures of atomic gases with reactive
collisions~\cite{vrh}, far-from-equilibrium quantum spin
chains~\cite{cas}, some models of Brownian
motors~\cite{SGN97,KMHLT06}, soft matter colloidal particles
\cite{ERAR10}, different nonequilibrium systems~\cite{royce}, and also
for the kinetically constrained models of glass
formers~\cite{JKGC08,S08,TPS12}.

Apart of these examples, negative differential mobility (NDM) has been
observed in the minimal model of a driven lattice gas, which captures
many essential features of the behavior in realistic systems.  In this
model one focuses on the dynamics of a hard-core tracer particle (TP)
which performs a random walk of mean waiting time $\tau$, biased by an
external force $F$, on a lattice containing a bath of hard-core
particles (or ``obstacles'') of density $\rho$, which perform
symmetric random walks of mean waiting time $\tau^*$.  Such a system
may be viewed as the combination of two paradigmatic models of
nonequilibrium statistical mechanics, namely the symmetric and
asymmetric exclusion processes, which have been extensively studied to
describe heat and particles transport properties \cite{CMZ11}.  Up to
now, only limiting situations of this model have been analyzed.

In the case of {\it immobile} bath particles ($\tau^*\to\infty$), it
has been argued that for a tracer subject to an external force and
diffusing on an infinite percolation cluster, the drift velocity
vanishes for large enough values of the force, and therefore NDM
occurs~\cite{barma}.  More recently, NDM was also observed via
numerical simulations for low density of immobile
particles~\cite{LF13,BBMS13} and analytically accounted for
~\cite{LF13}, but to the first order in $\rho$ only.  Surprisingly
enough, it appears that NDM is not a specific feature of a frozen
distribution of obstacles but also emerges in dynamical environments
undergoing continuous reshuffling due to obstacles random motion
($\tau^*<\infty$).  Indeed, very recently, numerical analysis
performed in~\cite{BM14} at a specific value of the density revealed
that NDM could occur in a 2D driven lattice gas for bath particles
diffusing slow enough.

In general, the origin of the NDM has been attributed to the
nonequilibrium (called ``frenetic'') contributions appearing in the
fluctuation-dissipation relation~\cite{LCZ05,BMW09}.  As shown earlier
in~\cite{BCCMO00,BCCMO01}, due to its interactions with the
environment the TP drives such a crowded system to a nonequilibrium
steady-state with a nonhomogeneous obstacles density profile.
However, the ``nonequilibrium'' condition is clearly not the only
necessary condition for the NDM to emerge - in simulations
in~\cite{BM14} this phenomenon is apparent for some range of
parameters but it definitely should be absent when the obstacles move
sufficiently fast so that the TP sees the environment as a fluid.

\begin{figure}[!t]
\includegraphics[width=.9\columnwidth,clip=true]{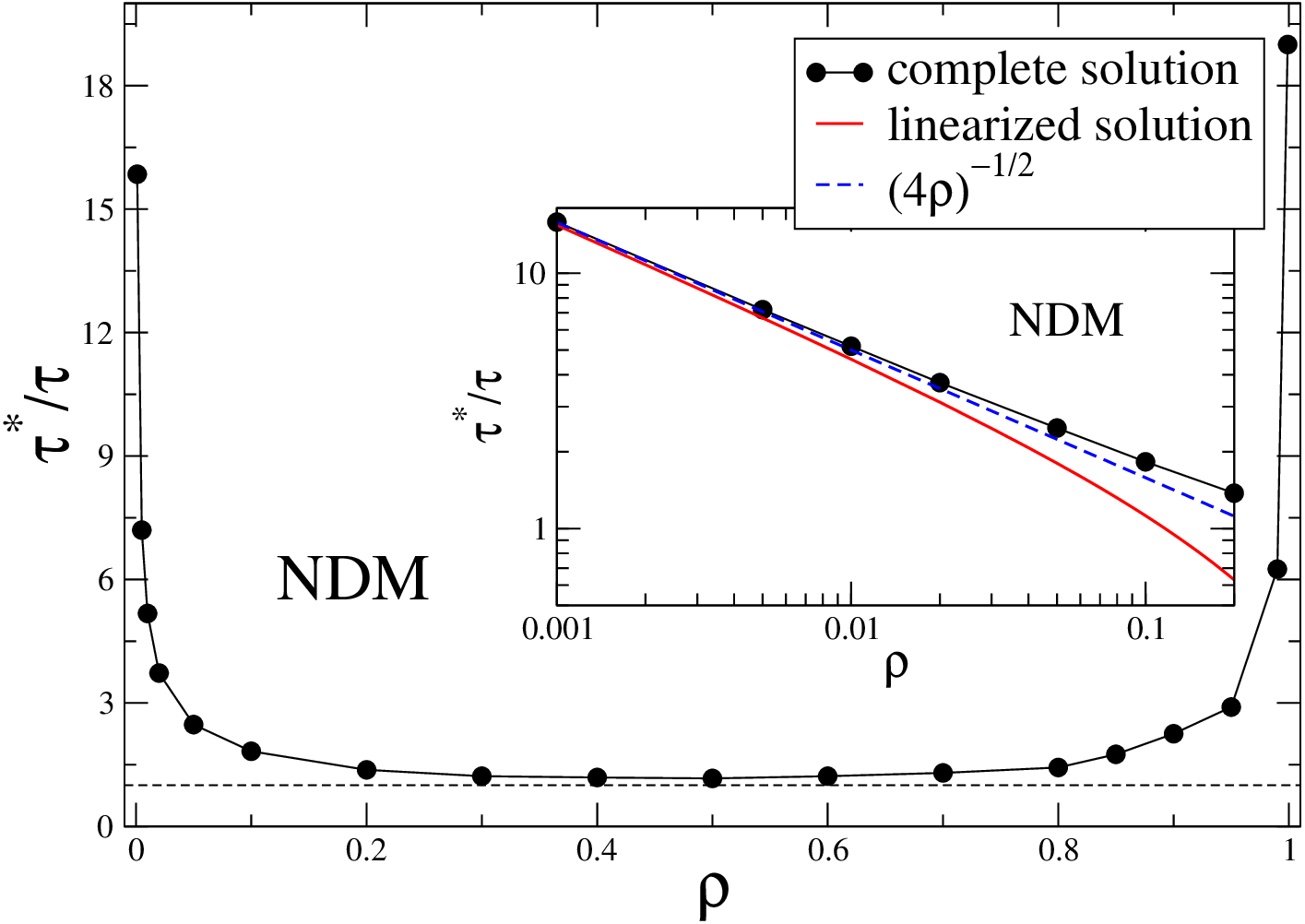}
\caption{(Color online) Region of negative differential mobility (NDM) in the plane $\tau^*/\tau$ vs
  $\rho$, for $d=2$ (black circles), as revealed by our analytical approach. Inset: zoom of the low density
  region and comparison with the prediction of the linear
  approximation, Eq.~(\ref{lin_system}) (red line), and with
  asymptotic result, Eq.~(\ref{asymp}) (blue dashed line).}
\label{fig1}
\end{figure}

Finally, NDM seems to be controlled by both the density $\rho$ and the
diffusion time scale $\tau^*$ of the bath particles. However, a
microscopic theoretical analysis of this effect is still lacking. The
only available analysis is restricted to the case of immobile
obstacles (in the low density regime) where, by definition, the bath
particles are not perturbed by the TP. In this Letter, we reveal the
complete scenario of this coupled dynamics providing i) a scaling
argument in the dilute regime that unveils the physical mechanism of
NDM, ii) an analytic analysis of the TP velocity for arbitrary values
of system parameters, and iii) a criterion for the NDM effect to be
observed, which shows in particular that for any $\rho$ NDM exists if
$\tau^*$ is large enough (see Fig.~\ref{fig1}).

More precisely, using a decoupling of relevant correlation functions,
we derive the force-velocity relation $V(F)$ valid for general $\rho$,
$\tau$, $\tau^*$ and $F$, and for any dimension $d\ge2$.  This
approximate expression is shown to be exact both in the dilute and in
the dense limit and provides results in excellent agreement with
numerical simulations for a wide range of parameters.  In the low
density regime, we recover the exact result obtained in~\cite{LF13} in
the limit $\rho\to 0$ and $\tau^* \to \infty$, while in the high
density limit our general expression gives back the exact results
of~\cite{BO02}.  Therefore, our theoretical framework unifies existing
asymptotic results~\cite{LF13,BBMS13,BM14}.  Our analytic result also
allows us to quantify the non-trivial nonmonotonic behavior of the
velocity with respect to the force, bringing to the fore the central
role of the coupling between density and time scales. In particular,
we analytically determine in the plane $(\rho,\tau^*/\tau)$ the region
for NDM and establish an accurate criterion for the existence of the
NDM (exact at linear order in $\rho$), see Fig.~\ref{fig1}~\cite{note2}.

\emph{Model.}--The dynamics in the system under study is defined as
follows. Each bath particle, selected at random, waits an exponential
time with mean $\tau^*$ and then selects the jump direction with
probability $1/2d$. Once the jump direction is chosen, the obstacle
attempts to move onto the target site: the move is realized if the
target site is empty at this time moment; otherwise, if the target
site is occupied by either another obstacle or the TP - the move is
rejected. In a similar fashion, the TP waits an exponential time with
mean $\tau$ and then chooses to jump in the direction $\nu$ ($\nu\in
\{\pm 1,\ldots,\pm d\}$) with probability
\begin{equation}
p_\nu=\frac{e^{(\beta/2)\boldsymbol{F}\cdot\boldsymbol{e}_\nu}}
{\sum_\mu e^{(\beta/2)\boldsymbol{F}\cdot\boldsymbol{e}_\mu}},
\label{pnu}
\end{equation}
where $\beta$ is the inverse temperature (measured in the units of the
Boltzmann constant), $\boldsymbol{e}_\mu$ are the corresponding $2d$
base vectors of the hypercubic lattice, the lattice step has been
taken equal to one and we denote $\boldsymbol{F}\equiv
F\boldsymbol{e}_1$.  Note that (\ref{pnu}) provides the standard
choice of the transition probabilities, which satisfy the generalized
detailed balance condition~\cite{LS99}, but arbitrary choices of
$p_\nu$~\cite{BM14} can be considered within our
formalism~\cite{note}.

Before discussing the mathematical details of our approach, we first
present a scaling argument that reveals the physical mechanism
underlying NDM and provides an estimation of the threshold in the low
density limit.  Assuming a strong external force, one has $p_1\simeq
1-\epsilon$, $p_{-1}=\mathcal{ O}(\epsilon^2)$ with
$\epsilon=2\exp(-\beta F/2)$, so that the mean velocity in the absence
of obstacles can be written $(1-\epsilon)/\tau$. The stationary
velocity in the presence of obstacles is then given by the mean
distance $1/\rho$ travelled by the TP between two obstacles divided by
the mean duration of this excursion, which is the sum of the mean time
of free motion $\tau/[\rho(1-\epsilon)]$ and of the mean trapping time
$\tau_{{\rm trap}}$ per obstacle. The escape from a trap results from
two alternative independent events: the TP steps in the transverse
direction (with rate $\epsilon/\tau$) or the obstacle steps away (with
rate $3/(4\tau^*)$, for $d=2$). This leads to $1/\tau_{{\rm
    trap}}=3/(4\tau^*)+\epsilon/\tau$, and finally
\begin{equation}
V(F)=\frac{1-\epsilon}{\tau+4\rho(1-\epsilon)\frac{\tau^*}{3+4\epsilon\tau^*/\tau}}.
\end{equation}   
From this formula, it can be viewed that $V$ is decreasing with $F$ at
large $F$ (i.e. small $\epsilon$), and therefore non monotonic with
$F$, as soon as $\tau^*\gtrsim\tau/\sqrt{\rho}$.  This unveils the
physical origin of NDM in the dilute regime, where two effects
compete. On the one hand a large force reduces the travel time between
two consecutive encounters with bath particles; on the other hand it
increases the escape time from traps created by surrounding
particles. Eventually, for $\tau^*$ large enough, such traps are
sufficiently long lived to slow down the TP when $F$ is increased.  In
order to get a rigorous and quantitative understanding of NDM for all
parameter values, we now analyze in detail the microscopic dynamics of
the model.

\emph{General expression of the velocity.}-- Let the Boolean variable
$\eta(\boldsymbol{R})=\{1,0\}$ denote the instantaneous occupation of
the site at position $\boldsymbol{R}$ by any of the obstacles,
$\eta\equiv \{\eta(\boldsymbol{R})\}$ denote the instantaneous
configuration of all such occupation variables and
$\boldsymbol{R}_{TP}$ - the instantaneous position of the driven
particle.  The stationary velocity $V(F)$ along the field direction is
easily shown to be given by (see Supplementary
  Material~\cite{SM})
\begin{equation}
V(F)\equiv \frac{d\langle \boldsymbol{R}_{TP}\cdot
\boldsymbol{e}_{1}\rangle}{dt}=\frac{1}{2d\tau^*}(A_1-A_{-1}),
\label{vel}
\end{equation}
where the coefficients $A_\nu$ ($\nu=\pm 1, \ldots, \pm d$) are
defined by the relation $A_\nu\equiv
1+\frac{2d\tau^*}{\tau}p_\nu(1-k(\boldsymbol{e}_{\nu}))$. Here,
$k(\boldsymbol{e}_{\nu})\equiv \sum_{\boldsymbol{R}_{TP},\eta}
\eta(\boldsymbol{R}_{TP}+\boldsymbol{e}_{\nu})P(\boldsymbol{R}_{TP},\eta)$
represents the stationary density profile around the TP,
$P(\boldsymbol{R}_{TP},\eta)$ being the joint probability
of finding the TP at the site $\boldsymbol{R}_{TP}$ with the configuration of obstacles
$\eta$.

In order to obtain a general  expression for the TP stationary
velocity for arbitrary force, we make use of the decoupling
approximation~\cite{BOMR96} for the correlation function of the
occupation variables of the form
\begin{equation}
\langle \eta(\boldsymbol{R}_{TP}+\boldsymbol{\lambda})\eta(\boldsymbol{R}_{TP}+\boldsymbol{e}_\nu)\rangle \approx
\langle\eta(\boldsymbol{R}_{TP}+\boldsymbol{\lambda}) \rangle \langle \eta(\boldsymbol{R}_{TP}+\boldsymbol{e}_\nu)\rangle,
\label{approx}
\end{equation}
which presumes that the occupation of the site just in front of the
TP, and of a site some distance $\lambda$ apart of it, become
statistically independent. This approach represents a mean-field-like
approximation and its physical motivation relies on the observation
that a fluctuation in the occupancy of the sites in the vicinity of
the tracer does not affect the dynamics far from the tracer
itself. This decoupling scheme has been previously used in
\cite{BCCMO00,BCCMO01} to derive general equations for the TP velocity
in two-dimensional open systems. However, the analysis in
\cite{BCCMO00,BCCMO01} has only been concerned with the linear
response regime, giving access to the Stokesian behavior of the
mobility and hence, via the Einstein relation, to the diffusion
coefficient of the particle in the absence of external bias. Here we
extend this analysis to nonlinear response (arbitrary force) and
arbitrary dimensionality of the embedding lattice in order to define
the physical conditions under which the NDM takes place.

Following~\cite{BCCMO01}, this decoupling approximation can be shown
to lead to a closed system for the $A_\nu$, which is reported
in~\cite{SM}.  This system is highly nonlinear in the coefficients
$A_\nu$. However, it can be numerically solved to find the analytic
value of the TP velocity for an arbitrary choice of the model
parameters.

\emph{Criterion for NDM.}--By using our analytical solution, the
region for NDM in the plane $(\rho,\tau^*/\tau)$ can be determined, as
reported in Fig.~\ref{fig1}, which constitutes the key result of this
Letter.  Importantly, this shows that for every density there exists a
value of $\tau^*/\tau$ above which NDM can be observed; this value
diverges for both $\rho\to0$ and $\rho\to1$. In turn, for any value of
$\tau^*/\tau\gtrsim1$, there exists a range of density
$[\rho_1,\rho_2]$, for which NDM occurs. When $\tau^*/\tau$ is
sufficiently large, the value of $\rho_1$ can be made explicit using a
small density expansion (see Eqs.(\ref{vel_lin})-(\ref{lin_system})
below). This leads to the exact asymptotic result
  \begin{equation}
\rho_1\underset{\tau^*/\tau\to \infty}{\sim} \frac{1}{4} \left(\frac{\tau}{\tau^*}\right)^2,
\label{asymp}
\end{equation}
which is validated numerically in Fig.~\ref{fig1}, see~\cite{SM}. Note
that this exact result is consistent with our earlier scaling
argument.

\begin{figure}[!tb]
\includegraphics[width=1\columnwidth,clip=true]{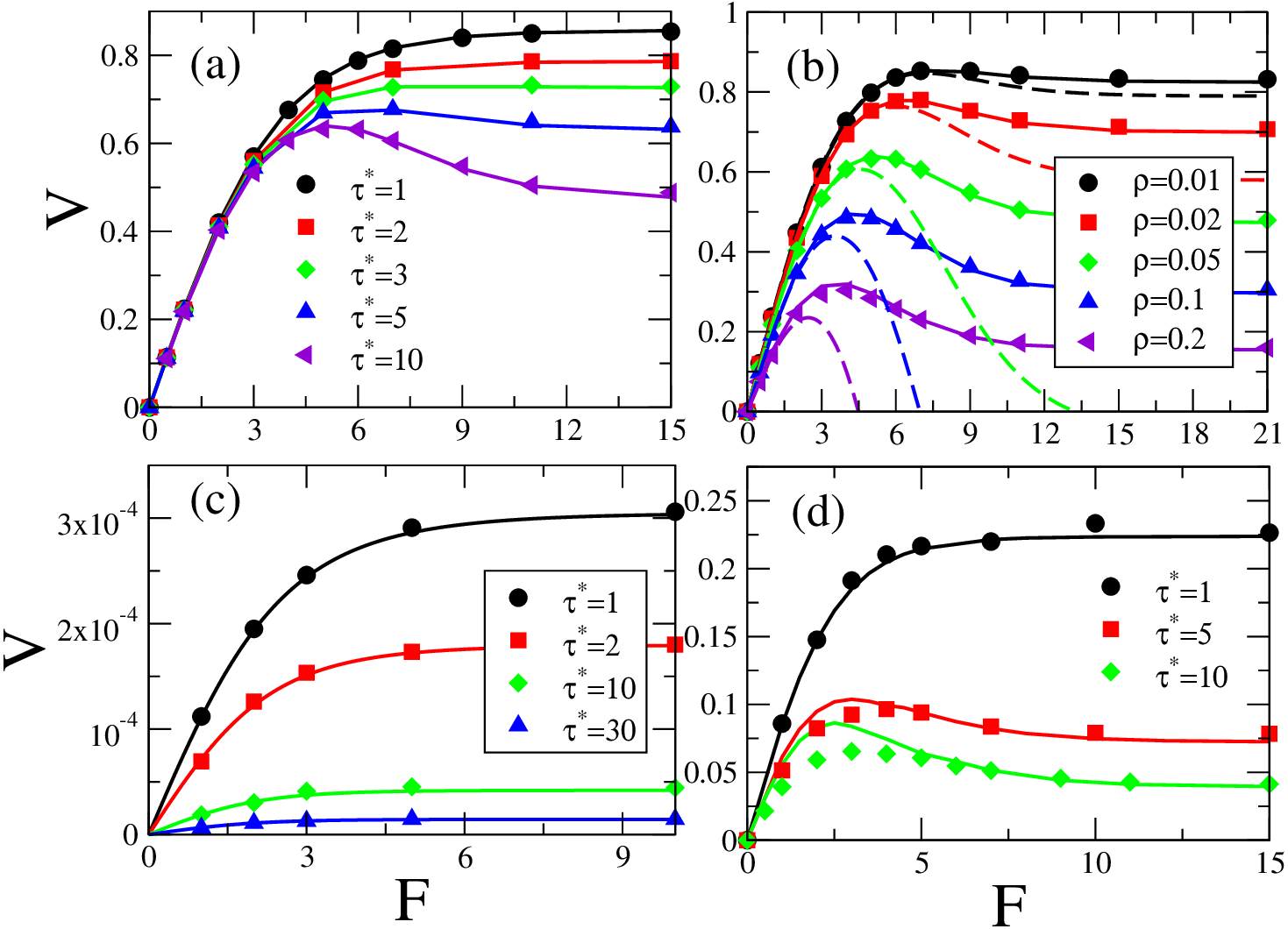}
\caption{(Color online). $V(F)$ for $d=2$ and $\beta=1$: (a) $\rho=0.05$, $\tau=1$ and
  different $\tau^*$, analytic prediction (lines) and numerical
  simulations (symbols); (b) $\tau=1$ and $\tau^*=10$, analytic
  prediction (continuous lines), numerical simulations (symbols) and
  linearized solution (dashed lines); (c) high density limit,
  $\rho=0.999$, with $\tau=1$ and different $\tau^*$, analytic
  prediction of Eq.~(\ref{rho1}) (lines) and numerical simulations
  (symbols); (d) $\rho=0.5$, $\tau=1$ and different $\tau^*$, analytic
  prediction (lines) and numerical simulations (symbols).}
\label{fig0}
\end{figure}

In order to validate the above scenario and to explore the
effectiveness of the decoupling approximation (\ref{approx}), we have
performed numerical simulations for different dimensions.  A very good
agreement is observed for a wide range of parameters (see
Fig.~\ref{fig0} for a two-dimensional infinite square lattice
and~\cite{SM} for the three-dimensional case). We show below that this
approximation is actually exact in both limits $\rho\to0$ (at linear
order in $\rho$) and $\rho\to1$.

\emph{Low density limit.} -- In the dilute limit $\rho\to 0$, the
system for the coefficients $A_\mu$ can be drastically simplified
(see~\cite{SM}). In this case, one has $A_\mu\sim
1+(2d\tau^*/\tau)p_\mu$ 
and the TP velocity can be expressed as
\begin{eqnarray}
V(\rho\to 0)&=&\frac{1}{\tau}(p_1-p_{-1}) \nonumber \\
&-&\frac{\rho}{\tau}(p_1-p_{-1}+p_1v_1-p_{-1}v_{-1})+o(\rho),\nonumber\\
\label{vel_lin}
\end{eqnarray}
where the coefficients $v_{\boldsymbol{n}}$ satisfy the linear system
of equations
\begin{eqnarray}
&&2d(1+\tau^*/\tau)v_{\boldsymbol{n}}=\sum_\nu[1+(2d\tau^*/\tau)p_\nu]v_{\boldsymbol{e}_\nu}\nabla_{-\nu}\mathcal{F}_{\boldsymbol{n}} \nonumber \\
&-&(2d\tau^*/\tau)(p_1-p_{-1})(\nabla_1-\nabla_{-1})\mathcal{F}_{\boldsymbol{n}}.
\label{lin_system}
\end{eqnarray}
Here, the functions $\mathcal{F}_{\boldsymbol{n}}$ depend on the
coefficients $A_\mu$, on dimension of the system, and are reported
explicitly in~\cite{SM}, while $\nabla_\mu$ is a differential operator
defined by the relation $\nabla_\mu f(\boldsymbol{\lambda})\equiv
f(\boldsymbol{\lambda}+\boldsymbol{e}_\mu)-f(\boldsymbol{\lambda})$.
The $V(F)$ obtained in this dilute limit is reported in
Fig.~\ref{fig0}(b) (dashed lines) for different densities and shows
the same behavior as the complete solution, even at intermediate
values of $\rho$, for small enough forces.

A further simplification occurs in the limit considered in \cite{LF13}
of the standard Lorentz gas, namely when $\tau^*/\tau\to\infty$.  In
this case, from~(\ref{lin_system}) we obtain an explicit solution,
which, as an actual fact, coincides with the analytic results
presented in~\cite{LF13}. In the particular case $d=2$, the functions
$\mathcal{F}_{\boldsymbol{n}}$ simplify to
\begin{equation}
\mathcal{F}_{\boldsymbol{n}}=e^{-n_1 F/2} \int_0^\infty e^{-t}\textrm{I}_{n_1}(2t/Z)
\textrm{I}_{n_2}(2t/Z)dt, 
\label{F_inf}
\end{equation}
with $Z=2+e^{\beta F/2}+e^{-\beta F/2}$.  Substituting
Eq.~(\ref{F_inf}) into the system~(\ref{lin_system}), and using
Eq.~(\ref{vel_lin}), one recovers the exact result of~\cite{LF13}
(see~\cite{SM}). As the accuracy of our analytic results increases
when $\tau^*/\tau$ decreases, as shown numerically in Fig.~\ref{fig0},
we claim that our decoupling approximation, Eq.~(\ref{approx}), is
exact at linear order in $\rho$.

\emph{High density limit.}--As detailed in~\cite{SM} and illustrated
here in the particular case $d=2$, the system for the coefficients $A_\mu$
linearized around $1-\rho$ leads to
  \begin{equation}
V(F)=\frac{1}{\tau}(1-\rho)\frac{\sinh(\beta F/2)}{1+\cosh(\beta F/2)[1+\frac{2\tau^*}{\tau}(\pi-2)]}.
\label{rho1}
\end{equation}
This result gives back the exact expression obtained in~\cite{BO02} in
the particular case $\tau=\tau^*$.

\emph{Conclusion.}--We have presented an analytic theory for NDM in a
general driven lattice gas. Exploiting a decoupling approximation, we
have obtained an analytic expression for the force velocity relation.
This expression which goes beyond linear response, is shown to be
exact in both $\rho\to0$ and $\rho\to1$ regimes and turns out to be in
very good agreement with numerical simulations for a wide range of
parameters. In particular, for values of $\tau^*$ large enough, a
nonmonotonic behavior of the TP velocity as a function of the external
force is indeed observed.  Our study extends analytical results
obtained in~\cite{LF13} and sheds light on recent numerical
observations~\cite{BBMS13,BM14}.  In particular, with the choice of
transition rates of~\cite{BM14}, which do not depend on the field in
the transverse direction, NDM is observed only for much larger values
of $\tau^*/\tau$. This is due to the fact that the escape time of the
TP from traps, in that case, is insensitive to the applied force to
linear order in $\rho$.

Our solution reveals and quantifies a minimal physical mechanism
responsible for NDM, which is based on the coupling between the
density of obstacles and the diffusion time scales of the TP and
obstacles. Our minimal model, which takes into account the repulsive
part of the particle-particle interactions only, suggests that the
phenomenon of the negative differential mobility should be a generic
feature of biased transport in crowded environments.

The work of O.B. and A.S. is supported by the European Research
Council (Grant No. FPTOpt-277998).

\onecolumngrid

\newpage

\begin{center}

\Large{\textbf{Supplemental Material}}

\end{center}

\section{Computation of the stationary velocity}

The time evolution of the joint probability
$P(\boldsymbol{R}_{TP},\eta; t)$ of finding at time $t$ the TP at the
site with the configuration of obstacles $\eta$, is governed by the
following master equation
\begin{eqnarray}
&&\partial_tP(\boldsymbol{R}_{TP},\eta; t)=\frac{1}{2d\tau^*} 
\sum_{\mu=1}^d \sum_{\boldsymbol{r}\ne \boldsymbol{R}_{TP}-\boldsymbol{e}_{\mu},\boldsymbol{R}_{TP}}[P(\boldsymbol{R}_{TP},\eta^{\boldsymbol{r},\mu}; t)-P(\boldsymbol{R}_{TP},\eta; t)]
\nonumber \\
&+&\frac{1}{\tau}\sum_{\mu=1}^dp_\mu\{[1-\eta(\boldsymbol{R}_{TP})]P(\boldsymbol{R}_{TP}-\boldsymbol{e}_{\mu},\eta; t)
-[1-\eta(\boldsymbol{R}_{TP}+\boldsymbol{e}_{\mu})]P(\boldsymbol{R}_{TP},\eta; t)\},
\end{eqnarray}
where $\eta^{\boldsymbol{r},\mu}$ is the configuration obtained from
$\eta$ by exchanging the occupation numbers of sites $\boldsymbol{r}$
and $\boldsymbol{r}+\boldsymbol{e}_\mu$.

The stationary velocity of the TP is obtained by multiplying both
sides of the master equation by $(\boldsymbol{R}_{TP} \cdot
\boldsymbol{e}_{1})$, summing over all possible configurations
$(\boldsymbol{R}_{TP},\eta)$, and taking the limit $t\to\infty$. This
yields the expression
\begin{equation}
V = \frac{1}{\tau}\left\{p_1\left[1-k(\boldsymbol{e}_{1})\right]-p_{-1} \left[1-k(\boldsymbol{e}_{-1})\right]\right\},
\end{equation}
where the functions $k({\lamb})$ are the stationary values (in the $t\to \infty$ limit) of $k({\lamb};t)$, defined as 
\begin{equation}
k({\lamb};t) =  \sum_{\boldsymbol{R}_{TP},\eta}
\eta(\boldsymbol{R}_{TP}+\lamb)P(\boldsymbol{R}_{TP},\eta;t).
\end{equation}
We define $k({\lamb};t)$ for $\lamb=\zz$ by $k(\zz;t)=0$. The evolution equations for $k(\lamb;t)$ may be obtained by multiplying the master equation by $\eta(\boldsymbol{R}_{TP}+\lamb)$ and summing over all the configurations of $(\boldsymbol{R}_{TP},\eta)$. We get the following equation:
\begin{equation}
2d\tau^*\partial_t k(\lamb;t)=\sum_\mu\left(\nabla_\mu-\delta_{\lamb,\emuu}\nabla_{-\mu}\right)k(\lamb;t)+\frac{2d\tau^*}{\tau}\sum_\nu p_\nu \moy{[1-\eta(\boldsymbol{R}_{TP}+\enu)]\nabla_\nu\eta(\boldsymbol{R}_{TP}+\lamb)},
\label{}
\end{equation}
where we introduced the average $\langle X(\boldsymbol{R})\rangle\equiv \sum_{\boldsymbol{R}_{TP},\eta}X(\boldsymbol{R})P(\boldsymbol{R}_{TP},\eta;t)$,
and $\nabla_\mu$ is a differential
operator defined by the relation $\nabla_\mu
f(\boldsymbol{\lambda})\equiv
f(\boldsymbol{\lambda}+\boldsymbol{e}_\mu)-f(\boldsymbol{\lambda})$.
In order to solve this equation, we use the decoupling approximation proposed in the main text:
\begin{equation}
\langle \eta(\boldsymbol{R}_{TP}+\boldsymbol{\lambda})\eta(\boldsymbol{R}_{TP}+\boldsymbol{e}_\nu)\rangle \approx
\langle\eta(\boldsymbol{R}_{TP}+\boldsymbol{\lambda}) \rangle \langle \eta(\boldsymbol{R}_{TP}+\boldsymbol{e}_\nu)\rangle,
\label{approx}
\end{equation} 
which is valid for $\lamb\neq\boldsymbol{e}_\nu$. For convenience, we also introduce the functions $h(\lamb;t)$, defined by
\begin{equation}
\label{ }
h(\lamb;t)\equiv k(\lamb;t)-\rho.
\end{equation}
One finally shows that $h({\lamb};t)$ satisfy the following evolution equations
\begin{eqnarray}
\label{hbulk}
2d\tau^*\partial_t h(\lamb;t) &=& \widetilde{L} h(\lamb;t) \quad \textrm{for} \quad {\boldsymbol \lambda} \notin \{{\bf 0},\pm\boldsymbol{e}_{1},\ldots,\pm\boldsymbol{e}_{d}\} \\
2d\tau^*\partial_t h(\lamb;t) &=& \widetilde{L} h(\lamb;t) + \rho(A_\nu - A_{-\nu}) \quad \textrm{for} \quad {\boldsymbol \lambda} \in \{{\bf 0},\pm\boldsymbol{e}_{1},\ldots,\pm\boldsymbol{e}_{d}\},
\label{hboundary}
\end{eqnarray}
with $\widetilde{L}\equiv\sum_\mu A_\mu\nabla_\mu$ and
$A_\mu=1+(2d\tau^*/\tau)p_\mu[\rho_0-h(\boldsymbol{e}_{\mu})]$. 


We introduce the auxiliary variable $\boldsymbol{\xi}=(\xi_1,\dots,\xi_d)$ and the generating function
\begin{equation}
H(\boldsymbol{\xi};t) = \sum_{n_1=-\infty}^\infty \sum_{n_2,\cdots,n_d=0}^{L-1} h_{n_1,\cdots,n_d}(t)\prod_{j=2}^d\xi_j^{n_j},
\label{generating_H}
\end{equation}
where the shorthand notation
$h(n_1\boldsymbol{e}_1+\dots+n_d\boldsymbol{e}_d;t)=h_{n_1,\dots,n_d}(t)$ has been used. $(n_1,\dots,n_d)$ are the components of the vector $\boldsymbol{n}$. If $(n_1,\dots,n_d)=\boldsymbol{e}_\nu$ then we use $h_{\boldsymbol{e}_\nu}\equiv h_\nu$.   
From Eqs.~(\ref{hbulk}) and~(\ref{hboundary}) we can show that $H(\boldsymbol{\xi};t)$ is the
solution of the following partial differential equation
\begin{equation}
2d\tau^* \partial_t H(\boldsymbol{\xi};t) = 
\left[ \frac{A_1}{\xi_1}+A_{-1}\xi_1 + A_2\sum_{j=2}^d\left(\frac{1}{\xi_j} + \xi_j \right)  - \alpha \right]H(\boldsymbol{\xi};t)+K(\boldsymbol{\xi};t),
\label{H}
\end{equation}
with $\alpha = A_1+A_{-1}+2(d-1)A_2$ and
\begin{eqnarray}
\label{defK}
K(\boldsymbol{\xi};t)&\equiv&A_1(\xi_1-1)h_{1}(t)+A_{-1}\left(\frac{1}{\xi_1}-1\right)h_{-1}(t)\nonumber\\
&+&A_2 \sum_{j=2}^d\left[ (\xi_j-1)h_j(t) +\left(\frac{1}{\xi_j}-1\right)h_{-j}(t) \right]+ \rho(A_1-A_{-1}) \left( \xi_1-\frac{1}{\xi_1}\right).
\end{eqnarray}
The stationary solution of Eq.~(\ref{H}) is
\begin{equation}
H(\boldsymbol{\xi})=\frac{K(\boldsymbol{\xi})}{\alpha} \frac{1}{1-\left[\frac{A_1}{\alpha} \frac{1}{\xi_1} + \frac{A_{-1}}{\alpha}\xi_1+\frac{A_2}{\alpha} \sum_{j=2}^d\left(\frac{1}{\xi_j}+\xi_j \right)  \right]}.
\end{equation}



We rewrite the auxiliary variables as $\xi_j=\ex{iq_j}$, and introduce the function
\begin{equation}
\mathcal{F}_{\boldsymbol{n}}=\frac{1}{(2\pi)^d}\int_{[-\pi,\pi]^d} dq_1\dots dq_d \frac{\prod_{j=1}^d \ex{-in_j q_j}}{1-\lambda(q_1,\dots,q_d)}
\label{def_Fn_inf}
\end{equation}
with
\begin{equation}
\lambda(q_1,\dots,q_d) = \frac{A_1}{\alpha} \ex{-iq_1} + \frac{A_{-1}}{\alpha} \ex{iq_1}+\frac{2A_2}{\alpha}\sum_{j=2}^d \cos q_j,
\end{equation}
so that $H(\boldsymbol{\xi})$ becomes
\begin{equation}
H(q_1,\dots,q_d) = \frac{K(q_1,\dots,q_d)}{\alpha}\frac{1}{1-\lambda(q_1,\dots,q_d)}.
\label{def_H_K_inf}
\end{equation}
Note that $\mathcal{F}_{\boldsymbol{n}}$ is the long-time limit of the generating
function of a biased random walk on $d$-dimensional lattice~\cite{H95}.  Using the definition of $\mathcal{F}_{\boldsymbol{n}}$ from Eq. (\ref{def_Fn_inf}), and taking the inverse Fourier transforms, we get
\begin{equation}
\frac{1}{1-\lambda(q_1,\dots,q_d)} = \sum_{n_1,\dots,n_d=-\infty}^\infty  \left( \prod_{j=1}^d e^{i\pi n_j q_j} \right)\mathcal{F}_{n_1,\dots,n_d}.
\end{equation}
Using Eq. (\ref{def_H_K_inf}),
\begin{equation}
H(q_1,\dots,q_d) = \frac{1}{\alpha}  \sum_{n_1,\dots,n_d=-\infty}^\infty   K(q_1,\dots,q_d) \mathcal{F}_{n_1,\dots,n_d}  \prod_{j=1}^d e^{i n_j q_j}.
\label{system_GF}
\end{equation}
Finally, using the definition of $K$ in Eq.~(\ref{defK}), writing $H(q_1,\dots,q_d)$ using Eq. (\ref{generating_H}) and identifying the terms from both sides of Eq. (\ref{system_GF}), one shows
that $h_{n_1,\dots,n_d}$ is given by the following system of $2d$ equations
\begin{equation}
\alpha h_{n_1,\dots,n_d} = \sum_\nu A_\nu h_\nu \nabla_{-\nu} \mathcal{F}_{n_1,\dots,n_d} - (1-\rho_0) (A_1 - A_{-1})(\nabla_1 -\nabla_{-1})\mathcal{F}_{n_1,\dots,n_d},
\label{system}
\end{equation}
where $(n_1,\dots, n_d)$ are taken equal to the coordinates of the base vectors $\{ \pm
\boldsymbol{e}_1,\dots, \pm \boldsymbol{e}_d\}$.  Noticing that $h_{\pm2}=\dots = h_{\pm d}$
for symmetry reasons, this system of $2d$ equations may be reduced to
a system of three equations ($\nu=\pm 1,2$)
\begin{equation}
A_\nu=1+\frac{2d\tau^*}{\tau}p_\nu\left[1-\rho-\rho(A_1-A_{-1})\frac{\textrm{det}C_\nu}{\textrm{det}C}\right].
\label{system_text}
\end{equation}
In the above expressions the matrix $C\equiv
(A_\mu\nabla_{-\mu}\mathcal{F}_{\boldsymbol{e}_\nu}-\alpha
\delta_{\mu,\nu})_{\mu,\nu}$,
$\alpha=\sum_\mu A_\mu$, and the matrix $C_\nu$ is obtained from the
matrix $C$ by replacing the column corresponding to the index $\nu$
with the column vector
$((\nabla_1-\nabla_{-1})\mathcal{F}_{\boldsymbol{e}_\nu})_\nu$.

Notice that the functions $\mathcal{F}_{\boldsymbol{n}}$, defined
in~(\ref{def_Fn_inf}) can be rewritten as
\begin{equation}
\label{open}
\mathcal{F}_{\boldsymbol{n}}=\left( \frac{A_{-1}}{A_1}\right)^{n_1/2} \int_0^\infty e^{-t}\textrm{I}_{n_1}(2\alpha^{-1}\sqrt{A_1 A_{-1}}t)
\prod_{i=2}^d \textrm{I}_{n_i}(2\alpha^{-1}A_2t)dt, 
\end{equation}
$n_i$ being the components of the base vector $\boldsymbol{n}$ and
$\textrm{I}_{i}(x)$ - the modified Bessel function of first kind.

\section{Linearized solution for low density}

For the general $d-$dimensional case the functions
$h_{\boldsymbol{n}}$ satisfy the system~(\ref{system}).  In order to
derive an approximated solution in the low density limit, we introduce
the variables $v_{\boldsymbol{n}}$ via the relation
\begin{equation}
h_{\boldsymbol{n}}=v_{\boldsymbol{n}}\rho.
\label{h_rho}
\end{equation}
When $\mathbf{n}=\mathbf{e_\nu}$, we use $v_{\boldsymbol{n}}=v_\nu$,
so that the expression for the tracer velocity becomes
\begin{equation}
V= \frac{1}{\tau}(p_1-p_{-1}) - \frac{\rho}{\tau} (p_1-p_{-1} + p_1v_1-p_{-1}v_{-1}).
\label{vel}
\end{equation}
In the low density limit $\rho\to 0$ the coefficients $A_\mu$ can be
approximated as
\begin{equation}
A_\mu\sim 1+2dxp_\mu,  \qquad \alpha=\sum_\mu A_\mu \sim 2d (1 + x),
\end{equation}
and, substituting the expression~(\ref{h_rho}) into~(\ref{system}),
one obtains the system satisfied by the variables $v_{\boldsymbol{n}}$
\begin{equation}
2d(1+x)v_{\boldsymbol{n}}
=\sum_{\nu=\pm 1,2}[1+2dxp_\nu]v_{\boldsymbol{e}_\nu}\nabla_{-\nu}\mathcal{F}_{\boldsymbol{n}} 
-2dx(p_1-p_{-1})(\nabla_1-\nabla_{-1})\mathcal{F}_{\boldsymbol{n}}, 
\label{lin_system}
\end{equation}
where $x\equiv \tau^*/\tau$.
Notice that the system~(\ref{lin_system}) obtained in the low density
approximation is linear in the variables $v_{\boldsymbol{n}}$.

\subsection{Lorentz lattice gas limit}

Let us consider the explicit case $d=2$, in the limit of the standard Lorentz
gas, namely when $x\to\infty$. Then
the functions $\mathcal{F}_{\boldsymbol{n}}$ simplify to
\begin{equation}
\mathcal{F}_{\boldsymbol{n}}=e^{-n_1 F/2} \int_0^\infty e^{-t}\textrm{I}_{n_1}(2t/Z)
\textrm{I}_{n_2}(2t/Z)dt, 
\label{F_inf}
\end{equation}
with $Z=2+e^{\beta F/2}+e^{-\beta F/2}$. Introducing the variables $u_i$, with $i=\pm 1,2$,
through the relation $v_i=(p_i-p_{-i})u_i$, we obtain the following linear system
\begin{eqnarray}
(p_1\nabla_{-1}\mathcal{F}_{\boldsymbol{e}_1}-1)u_1 + (p_{-1}\nabla_1\mathcal{F}_{\boldsymbol{e}_1})u_{-1}+(2p_2\nabla_2\mathcal{F}_{\boldsymbol{e}_1})u_2&=&(\nabla_1-\nabla_{-1})\mathcal{F}_{\boldsymbol{e}_1}, \nonumber \\
(p_1\nabla_{-1}\mathcal{F}_{\boldsymbol{e}_{-1}})u_1 + (p_{-1}\nabla_1\mathcal{F}_{\boldsymbol{e}_{-1}}-1)u_{-1}+(2p_2\nabla_2\mathcal{F}_{\boldsymbol{e}_{-1}})u_2&=&(\nabla_1-\nabla_{-1})\mathcal{F}_{\boldsymbol{e}_{-1}}, \nonumber \\
(p_1\nabla_{-1}\mathcal{F}_{\boldsymbol{e}_2})u_1 + (p_{-1}\nabla_1\mathcal{F}_{\boldsymbol{e}_2})u_{-1}+(p_2(\nabla_2+\nabla_{-2})\mathcal{F}_{\boldsymbol{e}_2})u_2&=&(\nabla_1-\nabla_{-1})\mathcal{F}_{\boldsymbol{e}_2}. 
\label{system_lorentz}
\end{eqnarray}
Notice that the expression~(\ref{F_inf}) corresponds to the perturbed
time evolution operator (integrated in time) introduced in
Ref.~\cite{LF13}. In order to explicitly recover the solution reported
in~\cite{LF13}, we notice that, using the expressions for the
probabilities $p_1=e^{F/2}/Z$, $p_{-1}=e^{-F/2}/Z$ and $p_2=1/Z$, the
following identities can be obtained
\begin{eqnarray}
\nabla_{-1}\mathcal{F}_{\boldsymbol{e}_1}=\mathcal{F}_0-e^{-F/2}\mathcal{F}_{\boldsymbol{e}_2} &\qquad& \nabla_1\mathcal{F}_{\boldsymbol{e}_1}=e^{-F}\mathcal{F}_{2\boldsymbol{e}_2}-e^{-F/2}\mathcal{F}_{\boldsymbol{e}_2} \nonumber \\
\nabla_{2}\mathcal{F}_{\boldsymbol{e}_1}=\mathcal{F}_{\boldsymbol{e}_1+\boldsymbol{e}_2}-e^{-F/2}\mathcal{F}_{\boldsymbol{e}_2} &\qquad& (\nabla_1-\nabla_{-1})\mathcal{F}_{\boldsymbol{e}_1}=e^{-F}\mathcal{F}_{2\boldsymbol{e}_2}-\mathcal{F}_0 \nonumber \\
\nabla_{-1}\mathcal{F}_{\boldsymbol{e}_{-1}}=e^F\mathcal{F}_{2\boldsymbol{e}_2}-e^{F/2}\mathcal{F}_{\boldsymbol{e}_2} &\qquad& \nabla_1\mathcal{F}_{-\boldsymbol{e}_1}=\mathcal{F}_0-e^{F/2}\mathcal{F}_{\boldsymbol{e}_2} \nonumber \\
\nabla_{2}\mathcal{F}_{\boldsymbol{e}_{-1}}=e^F\mathcal{F}_{\boldsymbol{e}_1+\boldsymbol{e}_2}-e^{F/2}\mathcal{F}_{\boldsymbol{e}_2} &\qquad& (\nabla_1-\nabla_{-1})\mathcal{F}_{\boldsymbol{e}_{-1}}=\mathcal{F}_0-e^{F}\mathcal{F}_{2\boldsymbol{e}_2} \nonumber \\
\nabla_{1}\mathcal{F}_{\boldsymbol{e}_{2}}=\mathcal{F}_{\boldsymbol{e}_1+\boldsymbol{e}_2}-\mathcal{F}_{\boldsymbol{e}_2} &\qquad& (\nabla_2-\nabla_{-2})\mathcal{F}_{\boldsymbol{e}_{2}}=\mathcal{F}_{2\boldsymbol{e}_2}+\mathcal{F}_0-2 \nonumber \mathcal{F}_{\boldsymbol{e}_2} \\
&&(\nabla_1-\nabla_{-1})\mathcal{F}_{\boldsymbol{e}_{2}}=(1-e^F)\mathcal{F}_{\boldsymbol{e}_1+\boldsymbol{e}_2}. \nonumber
\end{eqnarray}
Finally, expressing the functions $\mathcal{F}_{\boldsymbol{e}_{2}}$ and
$\mathcal{F}_{2\boldsymbol{e}_{2}}$ in terms of $\mathcal{F}_0$ and
$\mathcal{F}_{\boldsymbol{e}_1+\boldsymbol{e}_2}$, namely
\begin{eqnarray}
\mathcal{F}_{\boldsymbol{e}_{2}}=\frac{Z}{4}(\mathcal{F}_0-1), \qquad \mathcal{F}_{2\boldsymbol{e}_{2}}=\mathcal{F}_0\left(\frac{Z^2}{4}-1\right)-2e^{F/2}\mathcal{F}_{\boldsymbol{e}_1+\boldsymbol{e}_2}-\frac{Z^2}{4},
\end{eqnarray}
one can check that from the system~(\ref{system_lorentz}) the explicit
solution reported in~\cite{LF13} follows.

\subsection{Exact criterion for NDM in the low density limit}

The solution of the system~(\ref{lin_system}) gives the coefficients
$v_1$ and $v_{-1}$ appearing in the expression~(\ref{vel}). These
coefficients depend on $x$ and on the probabilities $\{p_\nu\}$,
$v_\mu=v_\mu(x, \{p_\nu\})$, both explicitly and implicitly through
the functions $\mathcal{F}_{\boldsymbol{n}}$. In order to find the condition for
negative differential mobility, we consider the case of large force,
such that
\begin{equation}
p_1=1-\epsilon \qquad p_{-1}=O(\epsilon^2) \qquad  p_{\mu\ne \pm1}=\frac{\epsilon}{2d-2}, 
\label{epsi}
\end{equation}
where $\epsilon$ is a small quantity. Substituting these expressions
into the definition of $\mathcal{F}_{\boldsymbol{n}}$, we can expand to the first
order in $\epsilon$ to get
\begin{equation}
\mathcal{F}_{\boldsymbol{n}}(x,\epsilon)=\mathcal{F}_{\boldsymbol{n}}^{(0)}(x)+\epsilon
\mathcal{F}_{\boldsymbol{n}}^{(1)}(x),
\label{F_epsi}
\end{equation}
where $\mathcal{F}_{\boldsymbol{n}}^{(0)}(x)=\mathcal{F}_{\boldsymbol{n}}(x,\epsilon=0)$
and $\mathcal{F}_{\boldsymbol{n}}^{(1)}(x)=\left .\frac{\partial}{\partial
    \epsilon} \mathcal{F}_{\boldsymbol{n}}(x,\epsilon)\right|_{\epsilon=0}$.
Next, substituting Eqs.~(\ref{epsi}) and~(\ref{F_epsi}) into the
solutions of the system~(\ref{lin_system}), and retaining only the
terms up to the order $\epsilon$, we obtain the expression for the
coefficients $v_\mu$
\begin{equation}
v_\mu(x,\epsilon)=v_\mu^{(0)}(x)+\epsilon v_\mu^{(1)}(x). 
\end{equation}
Notice that $v_\mu^{(0)}(x)$ and $v_\mu^{(1)}(x)$ still have both an
explicit and an implicit dependence on $x$, through the functions
$\mathcal{F}_{\boldsymbol{n}}^{(0)}(x)$ and $\mathcal{F}_{\boldsymbol{n}}^{(1)}(x)$. Thus,
for the tracer velocity~(\ref{vel}) to the order $\epsilon$ we have
\begin{eqnarray}
\tau V &=& 
1-\epsilon 
- \rho \left[1-\epsilon +(1-\epsilon)\left(v_1^{(0)}+\epsilon v_1^{(1)}\right)\right] \nonumber \\
&=& 1-\rho\left(1+v_1^{(0)}\right)
-\epsilon \left[1-\rho \left(1+v_1^{(0)}-v_1^{(1)}\right)\right].
\end{eqnarray}
Eventually, writing
\begin{equation}
V(x)=V^{(0)}(x)+\epsilon V^{(1)}(x),
\end{equation}
with
\begin{eqnarray}
V^{(0)}(x) &=& \frac{1}{\tau} \left[1-\rho \left(1+v_1^{(0)}(x)\right)\right] \\
V^{(1)}(x) &=& \frac{1}{\tau} \left[-1+\rho \left(1+v_1^{(0)}(x)-v_1^{(1)}(x)\right)\right], 
\end{eqnarray}
a general criterion for negative differential mobility can be
obtained by studying the sign of the term 
$V^{(1)}(x)$, which yields the condition
\begin{equation} 
1-\rho H(x) < 0,
\label{rhoHx}
\end{equation} 
where 
\begin{equation} 
H(x) = 1 + \left[v_1^{(0)}(x)-v_1^{(1)}(x)\right].
\label{Hx}
\end{equation} 

The functions $v_1^{(0)}(x)$ and $v_1^{(1)}(x)$ satisfy the system
obtained by expanding~(\ref{lin_system}) to the first order in
$\epsilon$. In particular, in the case $d=2$, to the zero order we
have
\begin{equation}
A_0V_0=B_0,
\end{equation}
where, dropping the dependence on $x$ in the functions $\mathcal{F}_{\boldsymbol{n}}$,
\begin{equation}
A_0=\left(
\begin{array}{ccc}
 \mathcal{F}^{(0)}_{0,0}+4 x (\mathcal{F}^{(0)}_{0,0}-\mathcal{F}^{(0)}_{1,0}-1)-\mathcal{F}^{(0)}_{1,0}-4 & \mathcal{F}^{(0)}_{2,0}-\mathcal{F}^{(0)}_{1,0} & 2 \mathcal{F}^{(0)}_{1,1}-2 \mathcal{F}^{(0)}_{1,0} \\
 (4 x+1) (\mathcal{F}^{(0)}_{-2,0}-\mathcal{F}^{(0)}_{-1,0}) & -4 x-\mathcal{F}^{(0)}_{-1,0}+\mathcal{F}^{(0)}_{0,0}-4 & 2 \mathcal{F}^{(0)}_{-1,1}-2 \mathcal{F}^{(0)}_{-1,0} \\
 (4 x+1) (\mathcal{F}^{(0)}_{-1,1}-\mathcal{F}^{(0)}_{0,1}) & \mathcal{F}^{(0)}_{1,2}-\mathcal{F}^{(0)}_{0,2} & -4 x+\mathcal{F}^{(0)}_{0,0}-2 \mathcal{F}^{(0)}_{0,1}+\mathcal{F}^{(0)}_{0,2}-4 \\
\end{array}
\right),
\end{equation}
\begin{equation}
V_0=\left(
\begin{array}{c}
 v_1^{(0)} \\
 v_{-1}^{(0)} \\
 v_2^{(0)} \\
\end{array}
\right),
\end{equation}
and
\begin{equation}
B_0=\left(
\begin{array}{c}
 4 x (\mathcal{F}^{(0)}_{2,0}-\mathcal{F}^{(0)}_{0,0}) \\
 4 x (\mathcal{F}^{(0)}_{0,0}-\mathcal{F}^{(0)}_{-2,0}) \\
 4 x (\mathcal{F}^{(0)}_{1,1}-\mathcal{F}^{(0)}_{-1,1}) \\
\end{array}
\right).
\end{equation}
Therefore
\begin{equation}
v_1^{(0)}=\frac{\text{det}\tilde{A_0}}{\text{det}A_0},
\end{equation}
where $\tilde{A_0}$ is obtained from $A_0$ replacing the first column
with the vector $B_0$.
Analogously, for $v_1^{(1)}$, we have to solve the system
\begin{equation}
A_1V_1=B_1,
\end{equation}
where
\begin{equation}
A_1=\left(
\begin{array}{ccc}
 \mathcal{F}^{(0)}_{0,0}+4 x (\mathcal{F}^{(0)}_{0,0}-\mathcal{F}^{(0)}_{1,0}-1)-\mathcal{F}^{(0)}_{1,0}-4 & \mathcal{F}^{(0)}_{2,0}-\mathcal{F}^{(0)}_{1,0} & 2 \mathcal{F}^{(0)}_{1,1}-2 \mathcal{F}^{(0)}_{1,0} \\
 (4 x+1) (\mathcal{F}^{(0)}_{-2,0}-\mathcal{F}^{(0)}_{-1,0}) & -4 x-\mathcal{F}^{(0)}_{-1,0}+\mathcal{F}^{(0)}_{0,0}-4 & 2 \mathcal{F}^{(0)}_{-1,1}-2 \mathcal{F}^{(0)}_{-1,0} \\
 (4 x+1) (\mathcal{F}^{(0)}_{-1,1}-\mathcal{F}^{(0)}_{0,1}) & \mathcal{F}^{(0)}_{1,2}-\mathcal{F}^{(0)}_{0,2} & -4 x+\mathcal{F}^{(0)}_{0,0}-2 \mathcal{F}^{(0)}_{0,1}+\mathcal{F}^{(0)}_{0,2}-4 \\
\end{array}
\right),
\end{equation}
\begin{equation}
V_1=\left(
\begin{array}{c}
 v_1^{(1)} \\
 v_{-1}^{(1)} \\
 v_2^{(1)} \\
\end{array}
\right),
\end{equation}
and
\begin{equation}
B_1=\left(
\begin{array}{c}
B_{11} \\
B_{12} \\
B_{13} \\
\end{array}
\right),
\end{equation}
with
\begin{eqnarray}
B_{11}&=&-v_1^{(0)} (\mathcal{F}^{(1)}_{0,0}-\mathcal{F}^{(1)}_{1,0})+2 v_{2}^{(0)} \mathcal{F}^{(1)}_{1,0}+v_{-1}^{(0)} \mathcal{F}^{(1)}_{1,0}
-2 v_{2}^{(0)} \mathcal{F}^{(1)}_{1,1}-v_{-1}^{(0)} \mathcal{F}^{(1)}_{2,0} \nonumber \\
&+&4 x
   \left[(v_1^{(0)}+1) \mathcal{F}^{(0)}_{0,0}+v_{2}^{(0)} \mathcal{F}^{(0)}_{1,0}-v_{2}^{(0)} \mathcal{F}^{(0)}_{1,1}-\mathcal{F}^{(0)}_{2,0}-\mathcal{F}^{(1)}_{0,0}-v_1^{(0)}
   (\mathcal{F}^{(0)}_{1,0}+\mathcal{F}^{(1)}_{0,0}-\mathcal{F}^{(1)}_{1,0})+\mathcal{F}^{(1)}_{2,0}\right], \\
B_{12}&=& -v_1^{(0)} (\mathcal{F}^{(1)}_{-2,0}-\mathcal{F}^{(1)}_{-1,0})+2 v_{2}^{(0)} \mathcal{F}^{(1)}_{-1,0}+v_{-1}^{(0)} \mathcal{F}^{(1)}_{-1,0} 
-2 v_{2}^{(0)} \mathcal{F}^{(1)}_{-1,1}-v_{-1}^{(0)} \mathcal{F}^{(1)}_{0,0} \nonumber \\
&+&4 x
   \left[(v_1^{(0)}+1) \mathcal{F}^{(0)}_{-2,0}+v_{2}^{(0)} \mathcal{F}^{(0)}_{-1,0}-v_{2}^{(0)} \mathcal{F}^{(0)}_{-1,1}-\mathcal{F}^{(0)}_{0,0}-\mathcal{F}^{(1)}_{-2,0}-v_1^{(0)}
   (\mathcal{F}^{(0)}_{-1,0}+\mathcal{F}^{(1)}_{-2,0}-\mathcal{F}^{(1)}_{-1,0})+\mathcal{F}^{(1)}_{0,0}\right], \\
B_{13}&=&-v_{2}^{(0)} \mathcal{F}^{(1)}_{0,0}-v_1^{(0)} (\mathcal{F}^{(1)}_{-1,1}-\mathcal{F}^{(1)}_{0,1})+2 v_{2}^{(0)} \mathcal{F}^{(1)}_{0,1}-v_{2}^{(0)} \mathcal{F}^{(1)}_{0,2}+v_{-1}^{(0)} \mathcal{F}^{(1)}_{0,2} \nonumber \\
&-&2 x \big[-2
   (v_1^{(0)}+1) \mathcal{F}^{(0)}_{-1,1}+v_{2}^{(0)} (\mathcal{F}^{(0)}_{0,0}-2 \mathcal{F}^{(0)}_{0,1}+\mathcal{F}^{(0)}_{0,2}) \nonumber \\
&+&2 (\mathcal{F}^{(0)}_{1,1}+\mathcal{F}^{(1)}_{-1,1}+v_1^{(0)}
   (\mathcal{F}^{(0)}_{0,1}+\mathcal{F}^{(1)}_{-1,1}-\mathcal{F}^{(1)}_{0,1})-\mathcal{F}^{(1)}_{1,1})\big]-v_{-1}^{(0)} \mathcal{F}^{(1)}_{1,2}.
\end{eqnarray}
Therefore
\begin{equation}
v_1^{(1)}=\frac{\text{det}\tilde{A_1}}{\text{det}A_1},
\end{equation}
where $\tilde{A_1}$ is obtained from $A_1$ replacing the first column with the vector $B_1$.

In order to obtain an explicit formula for $v_1^{(0)}(x)$ and $v_1^{(1)}(x)$ we expand the complete solutions in $1/x$. 
This provides the leading contribution of the function $H(x)$
for $x\to\infty$, which corresponds to the large $\tau^*/\tau$ limit.
First, we write the expressions for $\mathcal{F}_{\boldsymbol{n}}^{(0)}(x)$ and
$\mathcal{F}_{\boldsymbol{n}}^{(1)}(x)$:
\begin{eqnarray}
\mathcal{F}_{\boldsymbol{n}}^{(0)}(x)&=&(1 + 4 x)^{-n_1/2}\int_0^\infty~dt~e^{-t}
  \textrm{I}_{n_1}[t \sqrt{1 + 4 x}/(2 + 2 x)]\textrm{I}_{n_2}[ t/(2 + 2 x)], \\
\mathcal{F}_{\boldsymbol{n}}^{(1)}(x)&=&x \epsilon  (4 x+1)^{-n_1/2} \int_0^\infty~dt~e^{-t} \nonumber \\
&\times&\left[\textrm{I}_{n_1}\left(\frac{t \sqrt{4 x+1}}{2 x+2}\right) \left(2 n_2 \textrm{I}_{n_2}\left(\frac{t}{2 x+2}\right)+\frac{t
   \textrm{I}_{n_2+1}\left(\frac{t}{2 x+2}\right)}{x+1}\right)-\frac{t \textrm{I}_{n_1+1}\left(\frac{t \sqrt{4 x+1}}{2 x+2}\right) \textrm{I}_{n_2}\left(\frac{t}{2
   x+2}\right)}{(x+1) \sqrt{4 x+1}}\right].
\end{eqnarray}
Then, developing in $1/x$ these expressions up to the order $1/x^2$, we get
\begin{equation}
\mathcal{F}_{\boldsymbol{n}}^{(0)}(x)=\int_0^\infty~dt~e^{-t}\left[G^{(0,0)}_{\boldsymbol{n}}(t)
+\frac{1}{x}G^{_{(0,1)}}_{\boldsymbol{n}}(t)+\frac{1}{x^2}G^{(0,2)}_{\boldsymbol{n}}(t) + O\left(\frac{1}{x^3}\right)\right],
\label{F0x}
\end{equation}
and
\begin{equation}
\mathcal{F}_{\boldsymbol{n}}^{(1)}(x)=\int_0^\infty~dt~e^{-t}\left[G^{(1,0)}_{\boldsymbol{n}}(t)
+\frac{1}{x}G^{(1,1)}_{\boldsymbol{n}}(t)+\frac{1}{x^2}G^{(1,2)}_{\boldsymbol{n}}(t) + O\left(\frac{1}{x^3}\right)\right].
\label{F1x}
\end{equation}
Using the expansions~(\ref{F0x}) and~(\ref{F1x}) in the solutions for
$v_1^{(0)}$ and $v_1^{(1)}$ we get
\begin{eqnarray} 
v_1^{(0)}(x)&=&2x-\frac{7}{4x}+\frac{69}{64x^2} + o\left(\frac{1}{x}\right)^3, \\
v_1^{(1)}(x)&=&-4x^2-x+8-\frac{1}{8x}-\frac{553}{64x^2} + o\left(\frac{1}{x}\right)^3. \\
\end{eqnarray} 
Finally, using Eq.~(\ref{Hx}), we obtain
\begin{equation} 
H(x)= -7+\frac{311}{32 x^2}-\frac{13}{8x}+3x+4x^2 \underset{x\to\infty}{\sim} 4x^2,
\end{equation} 
and, thus, the condition on the density $\rho$, Eq.~(\ref{rhoHx}),
yields
\begin{equation} 
\rho \underset{x\to\infty}{\sim} \frac{1}{4x^2}.
\end{equation} 

\section{Explicit solution for high density}

Introducing the vacancy density $\rho_0\equiv 1-\rho$, and considering
the case $d=2$, for the coefficients $A_\nu$ one has
\begin{equation}
A_\nu=1+\frac{4\tau^*}{\tau}p_\nu(\rho_0-h_\nu),
\end{equation}
and
\begin{equation}
A_1-A_{-1}=\frac{4\tau^*}{\tau}[\rho_0(p_1-p_{-1})-p_1h_1+p_{-1}h_{-1}].
\label{AA1}
\end{equation}
In the high density limit, $\rho_0\to 0$, the system satisfied by the
functions $h_\nu$
\begin{eqnarray}
\alpha (h_1-h_{-1})&=&\sum_\nu A_\nu h_\nu\nabla_{-\nu}(\mathcal{F}_{1,0}-\mathcal{F}_{-1,0}) \nonumber \\
&-&(1-\rho_0)(A_1-A_{-1})(\nabla_1-\nabla_{-1})(\mathcal{F}_{1,0}-\mathcal{F}_{-1,0}),
\end{eqnarray}
can be linearized, yielding the solutions 
\begin{equation}
h_{\pm 1}=\mp\frac{\rho_0\frac{4\tau^*}{\tau}(p_1-p_{-1})(\mathcal{F}_{2,0}-\mathcal{F}_{0,0})}{4+(\mathcal{F}_{0,0}-\mathcal{F}_{0,2})\left[\frac{4\tau^*}{\tau}(p_1+p_{-1})-1\right]}.
\label{solh}
\end{equation}
Substituting Eq.~(\ref{solh}) into~(\ref{AA1}), and using the
definition $A_\nu\equiv
1+\frac{2d\tau^*}{\tau}p_\nu(1-k(\boldsymbol{e}_{\nu}))$, one obtains
the tracer velocity
\begin{equation}
V(\rho\to1)= \frac{1}{\tau}(p_1-p_{-1})\rho_0\frac{1}{1+\frac{4\tau^*}{\tau}\frac{(p_1+p_{-1})(4-8/\pi)}{8/\pi}},
\label{Vrho1}
\end{equation}
where we have used the result
$\mathcal{F}_{0,0}-\mathcal{F}_{2,0}=4-\frac{8}{\pi}$~\cite{H95}. Let us notice that
Eq.~(\ref{Vrho1}) is valid for a general choice of $p_\nu$. In particular,
using the definition of the probabilities in Eq.~(1) of the
main text, one immediately recovers the final result reported in
Eq.~(13) of the letter.


\section{Numerical simulations}

\begin{figure}[!ht]
\includegraphics[width=.3\columnwidth,clip=true]{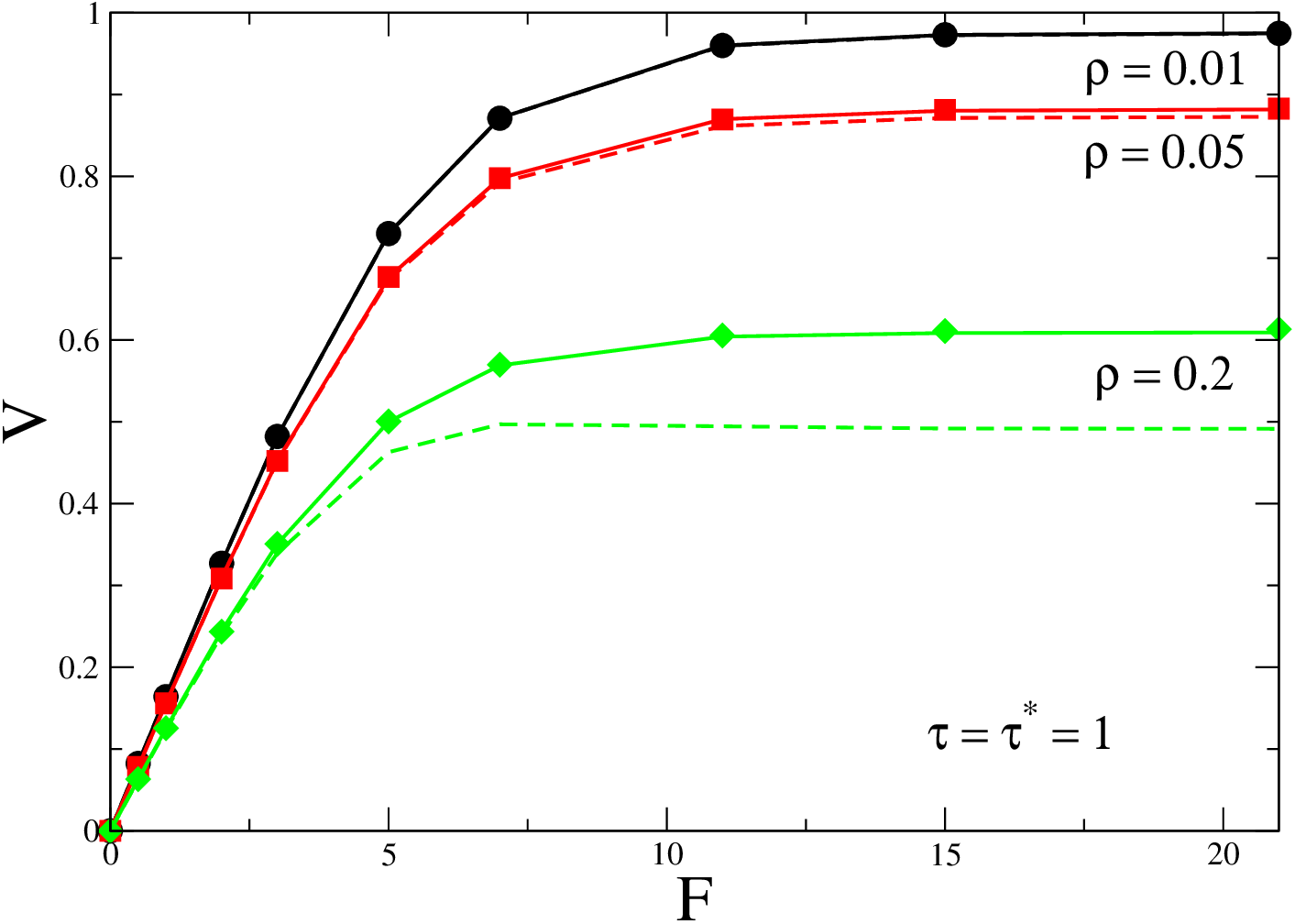}
\includegraphics[width=.3\columnwidth,clip=true]{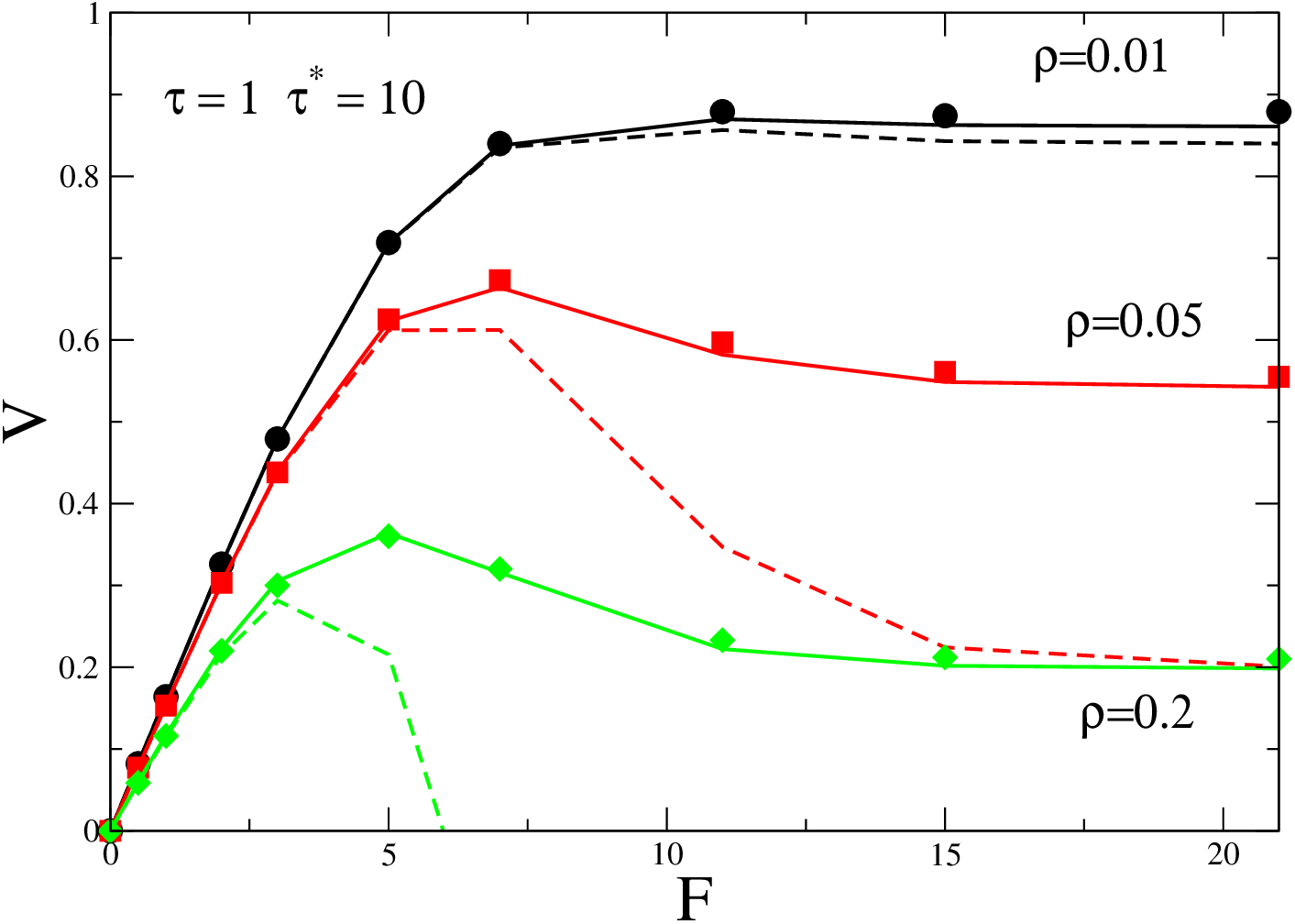}
\caption{3D lattice: Analytic prediction (continuous lines),
  numerical simulations (points) and explicit solution in the low
  density approximation (dashed lines) for the force velocity relation
  $V(F)$ in the case $d=3$, with $\tau=\tau^*=1$ (left panel) and
  $\tau=1$, $\tau^*=10$ (right panel).}
\label{fig3}
\end{figure}

We consider a $d-$dimensional lattice with $M$ sites and prepare the
$N$ particles in a random configuration, with density $\rho=N/M$.  In
the case $d=2$ for $\rho\leq 0.2$, we used a square lattice with
$M=L_x\times L_y=100^2$ sites, with periodic boundary conditions in
both directions, and we checked that results are independent of the
box size. In the case $\rho=0.5$, to avoid finite size effects, we
used $L_x=L_y=250$. 
For $d=3$, the box linear size is $L=60$, with periodic boundary
conditions.  In Fig~\ref{fig3} we compare analytic and numerical
results for the case $d=3$.


\begin{thebibliography}{99}

\bibitem{MPRV08}
U.~Marini Bettolo Marconi, A.~Puglisi, L.~Rondoni, and A.~Vulpiani, Phys. Rep. {\bf 461}, 111 (2008).

\bibitem{C11}
L.~Cugliandolo, J. Phys. A {\bf 44}, 483001 (2011).

\bibitem{seifert} 
U.~Seifert, Rep. Prog. Phys. {\bf 75}, 126001 (2012).

\bibitem{gradenigo2} G.~Gradenigo,  A.~Puglisi, A.~Sarracino, D.~Villamaina, and A.~Vulpiani, 
in {\em Nonequilibrium Statistical Physics of Small Systems: Fluctuation Relations and Beyond}, Eds.:
R.~Klages, W.~Just and C.~Jarzynski,  (Wiley-VCH, Weinheim, 2012)

\bibitem{M86}
A.~Morita, Phys. Rev. A {\bf 34}, 1499 (1986).

\bibitem{BB05}
J.-P.~Bouchaud and G.~Biroli, Phys. Rev. B {\bf 72}, 064204 (2005).

\bibitem{LCSZ08}
E.~Lippiello, F.~Corberi, A.~Sarracino, and M.~Zannetti, Phys. Rev. B {\bf 77}, 212201 (2008);
Phys. Rev. E {\bf 78}, 041120 (2008).

\bibitem{D12}
G.~Diezemann, Phys. Rev. E {\bf 85}, 051502 (2012).

\bibitem{ben1} O.~B\'enichou, P.~Illien, G.~Oshanin, and R.~Voituriez,
Phys. Rev. E {\bf 87}, 032164 (2013).

\bibitem{ben2} P.~Illien, O.~B\'enichou, C.~Mejia-Monasterio, G.~Oshanin, and R.~Voituriez,
Phys. Rev. Lett. {\bf 111}, 038102 (2013)

\bibitem{ben3}  O.~B\'enichou, A.~Bodrova, D.~Chakraborty, P.~Illien, A.~Law, C.~Mejia-Monasterio, G.~Oshanin, and R.~Voituriez, 
Phys. Rev. Lett. {\bf 111}, 260601 (2013)

\bibitem{weeks} P.~Habdas, D.~Schaar, A.~C.~Levitt, and E.~R.~Weeks, Europhys. Lett. {\bf 67}, 477 (2004).

\bibitem{SM09}
T.~M.~Squires, and T.~G.~Mason, Ann. Rev. Fluid Mech. {\bf 42}, 413 (2009).

\bibitem{poon} L.~G.~Wilson, A.~W.~Harrison, A.~B.~Schofield, J.~Arlt, and
W.~C.~K.~Poon, J. Phys. Chem. B {\bf 113}, 3806 (2009).

\bibitem{EM08}
D.~J.~Evans, and G.~Morriss. \emph{Statistical Mechanics of Non-equilibrium Liquids}, (Cambridge University Press, 2008).

\bibitem{WB09}
N.~J.~Wagner, and J.~F.~Brady, Physics Today {\bf 62}, 27 (2009).

\bibitem{royce} R.~K.~P.~Zia, E.~L.~Praestgaard, and O.~G.~Mouritsen, Am. J. Phys. {\bf 70}, 384 (2002). 

\bibitem{conwell} 
E.~Conwell, Physics Today {\bf 23}, 35 (1970).

\bibitem{NCCGO76}
F.~Nava, C.~Canali, F.~Catellani, G.~Gavioli, and G.~Ottaviani, J. Phys. C: Solid State Phys. {\bf 9}, 1685 (1976).

\bibitem{SBW86}
C.~J.~Stanton, H.~U.~Baranger, and J.~W.~Wilkins, Appl. Phys. Lett. {\bf 49}, 176 (1986).

\bibitem{LHC91}
X.~L.~Lei, N.~J.~M.~Horing, and H.~L.~Cui, Phys. Rev. Lett. {\bf 66}, 3277 (1991).

\bibitem{bryksin} H.~B\"ottger and V.~V.~Bryksin, Phys. Stat. Sol. (B) {\bf 113}, 9 (1982).

\bibitem{vrh} see, e.g., S.~B.~Vrhovac and Z.~Lj.~Petrovic, Phys. Rev. E {\bf 53}, 4012 (1996).

\bibitem{cas} G.~Benenti, G.~Casati, T.~Prosen, and D.~Rossini, Europhys. Lett. {\bf 85}, 37001 (2009).

\bibitem{SGN97}
G.~W.~Slater, H.~L.~Guo, and G.~I.~Nixon, Phys. Rev. Lett. {\bf 78}, 1170 (1997).

\bibitem{KMHLT06}
M.~Kostura, L.~Machura, P.~H\"anggi, J.~Luczka, and P.~Talkner, Physica A  {\bf 371}, 20 (2006).

\bibitem{ERAR10}
R.~Eichhorn, J.~Regtmeier, D.~Anselmetti, and P.~Reimann, Soft Matter {\bf 6}, 1858 (2010).

\bibitem{JKGC08} 
R.~L.~Jack, D.~Kelsey, J.~P.~Garrahan, and D.~Chandler, Phys. Rev. E {\bf 78}, 011506 (2008).

\bibitem{S08} 
M.~Sellitto, Phys. Rev. Lett. {\bf 101}, 048301 (2008).

\bibitem{TPS12} 
F.~Turci, E.~Pitard, and M.~Sellitto, Phys. Rev. E {\bf 86}, 031112 (2012).

\bibitem{CMZ11}
T.~Chou, K.~Mallick, R.~K.~P.~Zia, Rep. Prog. Phys. {\bf 74}, 116601 (2011).

\bibitem{barma} M.~Barma and D.~Dhar, J. Phys.: Solid State Phys.  {\bf 16},  1451 (1983).

\bibitem{LF13} 
S.~Leitmann and T.~Franosch, Phys. Rev. Lett. {\bf 111}, 190603 (2013).

\bibitem{BBMS13} 
P.~Baerts, U.~Basu, C.~Maes, and S.~Safaverdi, Phys. Rev. E {\bf 88}, 052109 (2013).

\bibitem{BM14} 
U.~Basu and C.~Maes, J. Phys. A: Math. Theor. {\bf 47}, 255003 (2014).

\bibitem{LCZ05}
E.~Lippiello, F.~Corberi, and M.~Zannetti,
Phys. Rev. E {\bf 71}, 036104 (2005).

\bibitem{BMW09}
M.~Baiesi, C.~Maes, and B.~Wynants,
Phys. Rev. Lett. {\bf 103}, 010602 (2009).

\bibitem{BCCMO00}
O.~B\'enichou, A.~M.~Cazabat, J.~De Coninck, M.~Moreau, and G.éOshanin, 
Phys. Rev. Lett. {\bf 84}, 511 (2000).

\bibitem{BCCMO01} 
O.~B\'enichou, A.~M.~Cazabat, J.~De Coninck, M.~Moreau, and G.~Oshanin, Phys. Rev. B {\bf 63}, 235413 (2001).

\bibitem{BO02}
O.~B\'enichou, and G.~Oshanin, Phys. Rev. E {\bf 66}, 031101 (2002).

\bibitem{note2} Let us notice that the change of behavior observed in
  the model does not represent a genuine phase transition occurring in
  the system.

\bibitem{LS99}
J.~L.~Lebowitz and H.~Spohn, J. Stat. Phys {\bf 95}, 333 (1999).

\bibitem{note} The choice of transition rates studied in~\cite{BM14}
  corresponds to take, for $d=2$, $p_1=(1/2)e^{\beta F/2}/(e^{\beta
    F/2}+e^{-\beta F/2}), p_{-1}=(1/2)e^{-\beta F/2}/(e^{\beta
    F/2}+e^{-\beta F/2})$ and $p_2=p_{-2}=1/4$, with $\tau=1/2$ and
  $\tau^*=1/4\gamma$, where $\gamma$ is the inverse time-scale
  introduced in~\cite{BM14}.

\bibitem{SM}
See Supplemental Material [url], which includes Ref.~\cite{hughes}, 
for details on the calculations and numerical simulations.

\bibitem{hughes}
B.~D.~Hughes, \emph{Random Walks and Random Environments} (Oxford Science, Oxford, 1995).

\bibitem{BOMR96}
S.~F.~Burlatsky, G.~Oshanin, M.~Moreau, and W.~P.~Reinhardt, Phys. Rev. E {\bf 54}, 3165 (1996).





\end{thebibliography}

\begin{thebibliography}{99}


\bibitem{H95}
B.~D.~Hughes, \emph{Random Walks and Random Environments} (Oxford Science, Oxford, 1995).

\bibitem{LF13}
S.~Leitmann and T.~Franosch, Phys. Rev. Lett. {\bf 111}, 190603 (2013).

\end{thebibliography}
\end{document}